\documentclass[twocolumn,aps,prb,showpacs]{revtex4}
\usepackage{epsfig}

\begin{document}


\title{Electron correlation and impurity-induced quasiparticle resonance states in cuprate superconductors}
\author{Bin Liu}
\email{liubin@bjtu.edu.cn}
\affiliation{Department of Physics, Beijing Jiaotong University, Beijing 100044, China}

\author{Xu Yan, and Feng Yuan}
\affiliation{Department of Physics, Qingdao University, Qingdao 266071, China}

\begin{abstract}

We theoretically study the quasiparticle resonance states around a nonmagnetic impurity in cuprate superconductors based on t-t'-J-U model. The purpose of introducing the Coulomb repulsive interaction U is to partially impose the double occupancy constraint by employing the Gutzwiller projected mean-field approximation. We determine the spatial variation of the order parameter and the local density of states (LDOS) by self-consistently solving Bogoliubov-de Gennes equations. We find that in the large U limit, a zero-energy resonance peak in the LDOS indeed appears for the impurity potential in the unitary limit, at the same time the asymmetric superconducting coherence peaks are strongly suppressed. As U decreases the electron double occupancy d is permitted and gradually increases, leading to the decreasing of order parameter. In particular the above zero-energy resonance peak begins to evolve into a double-peaked structure since a critical value $d_{c}$. These important feathers are qualitatively agreement with the scanning tunneling spectroscopy experiments, and uncover the essential role played by the electron correlation in cuprate superconductors.

\end{abstract}

\pacs{74.25.Jb, 71.10.Fd, 74.72.-h, 74.25.Ha}

\maketitle

\section{ Introduction}

After nearly two decades intense study of the anomalous properties of high-Tc superconductors (HTS), many important questions still remain open. Among others, the nonmagnetic impurities effect on the cuprates has always attracted much attention \cite{Lee,Hirschfeld,Balatsky0,Salkola,Franz,Yazdani,Zhu,Tsuchiura,Balatsky,pan,pan1,Morr,wang,Terashima,Balatsky1,bin,feng,Allou,Zeljkovic,Parham}. As we know, in the conventional s-wave superconductors, nonmagnetic impurities have little effect on the superfluid density and the transition temperature, which can be understood well from Anderson theorem\cite{Anderson}. However, for HTS, it is found that such impurities can cause a strong pair-breaking effect\cite{Ueda}, implying that HTS have the unconventional, most likely d-wave pairing symmetry\cite{Scalapino}. So understanding of the effects of the nonmagnetic impurities on these materials provide us important information to understand the pairing mechanism in HTS.

Experimentally, scanning tunneling microscopy (STM) is an ideal technique
for the study of such effects at the atomic scale\cite{Yazdani,pan,pan1,Terashima,Zeljkovic,Parham}. With the help of the high quality of the surface properties of the samples and the improvement of the experimental techniques, a great deal of reliable data have been
obtained by different STM groups. It is found that away from the nonmagnetic impurity, the tunneling spectra
show the typical asymmetric superconducting coherence peaks, and around the nonmagnetic impurity,
strong intra-gap density of states peaks are induced at energies close to the Fermi level, and at the same time
the superconducting coherence peaks are strongly suppressed.

Theoretically, it has been widely accepted that the essential physics of cuprates can be effectively described
by the two-dimensional Hubbard model or its equivalent t-J model in the large U limit. Based on these models, especially the t-J-like models, a good deal of work has been carried out to study the nonmagnetic impurity effect on the cuprates\cite{Balatsky0,Franz,Zhu,Morr,wang,Balatsky1}. But as we know that in the t-J-like models, the virtual double occupancy is completely neglected, and this maybe leads not to describe the real detailed physics in these materials. In this case, Laughlin proposed a new idea ``Gossamer superconductor" to describe the physics of the HTS \cite{Laughlin}. In the Gossamer superconductor, even for the half filling, it may be a superconductor because of the double occupancy. Stimulated by the Laughlin's idea, Anderson and
Ong\cite{Anderson0} proposed a new wave function to quantitatively explain the observed asymmetric tunneling conductivity in
the STM within the Gutzwiller-Resonating valence Bond theory \cite{Gutzwiller,Anderson1}. They believed that the asymmetries are predicted
not to exist within the Fermi liquid theory, and one needs to explain the STM results within the Gutzwiller projected
mean-field-theory.

Following above ideas, in this paper, we study the
nonmagnetic impurity effect in the cuprate superconductor within the two-dimensional t-t'-J-U model using
the Gutzwiller-projected mean-field-theory (MFT) and the Bogoliubov-de Gennes theory\cite{Zhang,Yuan}. The order parameter (OP) are determined self-consistently and the LDOS is calculated numerically. We reproduced the main experimental results
within our present theory. In the large U limit without electron double occupancy (EDO), far away from the local nonmagnetic impurity the LDOS shows asymmetric superconducting coherence peaks. While around the nonmagnetic impurity, a zero-energy resonance peak in the LDOS indeed appears when pushing the impurity potential into the unitary limit, and meanwhile the superconducting coherence peaks are
strongly suppressed. We also find that with increasing the EDO d which is directly
modulated by Coulomb repulsion U, the OP gradually decreases and the resulting superconducting coherence peaks move to lower energies, while it is interesting to see the above zero-energy resonance peak begins to evolve into a double-peaked structure since a critical value $d_{c}$. These novel feathers of asymmetric or splitting of the resonance state near Fermi energy are qualitatively agreement with the STM experiments\cite{Yazdani,pan}, and reveal the essential role played by the electron correlation in cuprate superconductors.

\section{the $t$-$t'$-$J$-$U$ Model and Gutzwiller projected mean-field approximation}

We start from the t-t'-J-U model on a square lattice \cite{Zhang,Yuan},
\begin{eqnarray}
 H&=&-t\sum_{i\hat{\eta}\sigma}C_{i \sigma}^\dagger C_{i+\hat{\eta}\sigma}+t'\sum_{i\hat{\tau}\sigma}C_{i \sigma}^\dagger C_{i+\hat{\tau}\sigma}\nonumber \\&+&J\sum_{i\hat{\eta}}S_i\cdot S_{i+\hat{\eta}}-\mu\sum_{i\sigma}C_{i \sigma}^\dagger C_{i \sigma}
 \nonumber \\&+&U\sum_i \hat{n}_{i\uparrow} \cdot \hat{n}_{i\downarrow}+\sum_{i\sigma}U_{i}n_{i\sigma}
\end{eqnarray}
where $\hat{\eta}=\pm\hat{x}$ and $\pm\hat{y}$, $\hat{\tau}=\pm\hat{x}\pm\hat{y}$, $C_{i \sigma}^\dagger$($C_{i \sigma}$) is the electron creation(annihilation) operator, $S_i=\frac{1}{2}\sum_{\sigma \sigma'} C_{i \sigma}^\dagger \overrightarrow{\sigma}_{\sigma \sigma'} C_{i \sigma}$ is spin operator with $\overrightarrow{\sigma}=({\sigma_x,\sigma_y,\sigma_z})$ as the Pauli matrices, $n_{i\sigma}=C_{i \sigma}^\dagger C_{i \sigma}$, $\mu$ is the chemical potential, and $U$ is the on-site Coulomb potential, which is introduced to partially impose the no-double-occupancy constraint for the strongly correlated system. In the limit U$\rightarrow\infty$, the model is reduced to the t-J model. The scattering potential from the single-site impurity is modeled by
$U_i=U_0\delta_{iI}$ with I the index for the impurity site.

To study the Hamiltonian (1) with the Gutzwiller variational approach, we take the trial wave function $|\psi\rangle$ as
\begin{eqnarray}
|\psi\rangle=P_{G}|BCS(\Delta_{ij})\rangle,
\end{eqnarray}
where $|BCS(\Delta_{ij})\rangle$ is the BCS mean-field solution, and $P_{G}$ is the Gutzwiller projection operator which is
defined as
\begin{eqnarray}
P_{G}=\Pi_{i}[1-(1-g)\hat{n}_{i\uparrow}\hat{n}_{i\downarrow}],
\end{eqnarray}
here $g$ is a variational parameter which takes the value between 0 and 1. The choice $g=0$ corresponds to the situation with no
doubly occupied sites($U\rightarrow\infty$), while $g=1$ corresponds to the uncorrelated state($U=0$). With the help of the trial wave function and the Gutzwiller approximation \cite{Gutzwiller}, we can get a Gutzwiller renormalized hamiltonian\cite{Yuan},
\begin{eqnarray}
 H_{eff}&=&-g_t(t\sum_{i\hat{\eta}\sigma}C_{i \sigma}^\dagger C_{i+\hat{\eta}\sigma}-t'\sum_{i\hat{\tau}\sigma}C_{i \sigma}^\dagger C_{i+\hat{\tau}\sigma})\nonumber \\
 &+&g_sJ\sum_{i\hat{\eta}}S_i\cdot S_{i+\hat{\eta}}+NUd\nonumber \\
 &+&\sum_{i\sigma}U_{i}n_{i\sigma}-\mu\sum_{i\sigma}C_{i \sigma}^\dagger C_{i \sigma}
\end{eqnarray}
where $g_t$ and $g_s$ are the renormalized factors in the Gutzwiller approximation,
\begin{eqnarray}
 g_t&=&\frac{2(n_{i}-2d_{i})}{n_{i}(2-n_{i})}\left[\sqrt{1-n_{i}+d_{i}}+\sqrt{d_{i}}\right]^{2}
\end{eqnarray}
\begin{eqnarray}
 g_s&=&\left[\frac{2(n_{i}-2d_{i})}{n_{i}(2-n_{i})}\right]^{2}
\end{eqnarray}
with the electron number $n_{i}$, and the double occupancy number $d_{i}$ at the site i. Then using the mean-field approximation, we obtain a Bogoliubov-de Gennes (BdG) equation,
\begin{eqnarray}
\sum_j \left( \begin{array}{cc}
 H_{ij} & F_{ij} \\ F^{*}_{ij} & -H^{*}_{i j}
\end{array}
\right) \left( \begin{array}{c}
u^{n}_{j}\\ v^{n}_{j}
\end{array}
\right) =E_n \left( \begin{array}{c}
u^{n}_{j}\\ v^{n}_{j}
\end{array}
\right),
\end{eqnarray}
with
\begin{eqnarray}
H_{ij}&=&-\sum_{\hat{\eta}}(g_tt+\frac{3}{4}g_sJ\chi_{ij})\delta_{j,{i+\hat{\eta}}}+\sum_{\hat{\tau}}g_tt'\delta_{j,{i+\hat{\tau}}}\nonumber \\
 &+&(U_i-\mu)\delta_{ij}\nonumber\\
F_{ij}&=&-\sum_{\hat{\eta}}\frac{3}{8}g_sJ\Delta_{ij}\delta_{j,{i+\hat{\eta}}}
\end{eqnarray}
In the above equations, we have introduced the electron pairing OP and the bond OP,
\begin{eqnarray}
\Delta_{ij} &=& \langle
C_{i\downarrow}C_{j\uparrow}-C_{i\uparrow}C_{j\downarrow}\rangle_{0}\nonumber\\
\chi_{ij}&=&\langle C^{\dag}_{i\uparrow}C_{j\uparrow}\rangle_{0}
\end{eqnarray}
which can be determined self-consistently as,
\begin{eqnarray}
\Delta_{ij}&=&\frac{1}{2}\sum_{n}(v_j^{n\ast}u_i^n+v_i^{n\ast}u_j^n)\tanh(\frac{1}{2}\beta E_n)\nonumber\\
\chi_{ij}&=&\sum_{n}u_i^nu_j^{n\ast}f(E_n)+\sum_{n}v_i^nv_j^{n\ast}[1-f(E_n)]\nonumber\\
n_{i}&=&\sum_{n}|u_i^n|^2f(E_n)+\sum_{n}|v_i^n|^2[1-f(E_n)]
\end{eqnarray}
and
\begin{eqnarray}
 0&=&\sum_{i,n}(\frac{\partial E_{i,n}}{\partial g_t}\frac{\partial g_t}{\partial d_{i}}+\frac{\partial E_{i,n}}{\partial g_s}\frac{\partial g_s}{\partial d_{i}})+U,
\end{eqnarray}
where $f(E)=1/(e^{\beta E}+1)$ is the Fermi-Dirac distribution function. In the numerical calculation, we construct a superlattice with the square lattice $Nx \times Ny$ as a unit supercell. As detailed in Ref. 7, this method can provide the required energy resolution for the possible resonant states. Throughout this paper,
we take the size of the unit supercell $N = 33\times33$, the number of supercell $Nc = 10\times10$. Then we can solve numerically the BdG equation and carry out an iteration until the selfconsistent
equations are satisfied. Hereafter, we set $t=1$, $t'/t=-0.3$, $J/t=0.3$ for the band structure corresponding to the doping $\delta= 0.1$. The impurity potential $U_{0}= 100t$ is in the unitary limit.

\section{Numerical Results and Discussions}

\begin{figure}[tbp]
\begin{center}
\includegraphics[width=0.8\linewidth]{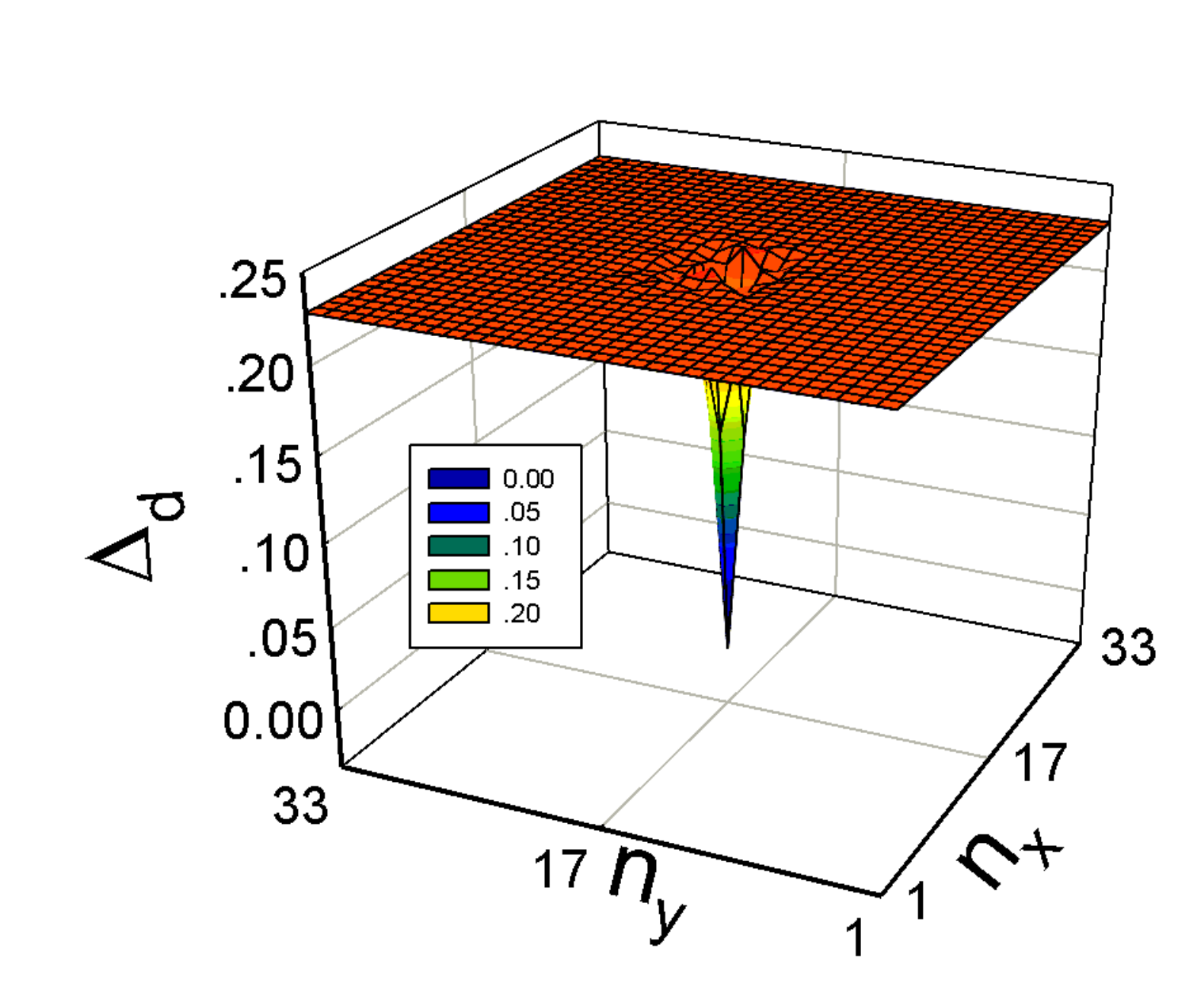}
\end{center}
\caption{(Color online) The spatial variation of the d-wave order parameter $\Delta_{d}$ for the parameter $U_{0}=100t$ in the large U limit.}
\end{figure}

We firstly review the local electronic structure near a nonmagnetic impurity in the limit U$\rightarrow\infty$, the model now is reduced to the t-t'-J model and no EDO is constraint. In Fig. 1, we plot the obtained OP. The spatial variation of the d-wave OP defined as \cite{Zhu}
\begin{eqnarray}
\Delta_{d}(i)=\frac{1}{4}[\Delta_{(i,i+x)}+\Delta_{(i,i-x)}\nonumber\\-\Delta_{(i,i+y)}-\Delta_{(i,i-y)}],
\end{eqnarray}
It is shown that because of the presence of the nonmagnetic impurity, the OP is
suppressed at the impurity site and recovers its bulk value over 2-3 lattice spacings.

Next we calculate the LDOS as,
\begin{eqnarray}
\rho_{i}(E)&=&-2\sum_{n,{\bf k}}[|u_i^{n,{\bf k}}|^2f'(E_{n,{\bf k}}-E)\nonumber \\
&+&|v_i^{n,{\bf k}}|^2f'(E_{n,{\bf k}}+E)]
\end{eqnarray}
where the prefactor 2 comes from the spin summation, and $f'(E) = df(E)/dE$ is the derivation of the fermi distribution
function $f(E)$. The LDOS $\rho_{i}$(E) is proportional to the local differential tunneling conductance which can be measured
in a scanning tunneling microscope/spectroscopy experiment, so we can compare our calculated LDOS with the STM results directly.

\begin{figure}[tbp]
\begin{center}
\includegraphics[width=0.9\linewidth]{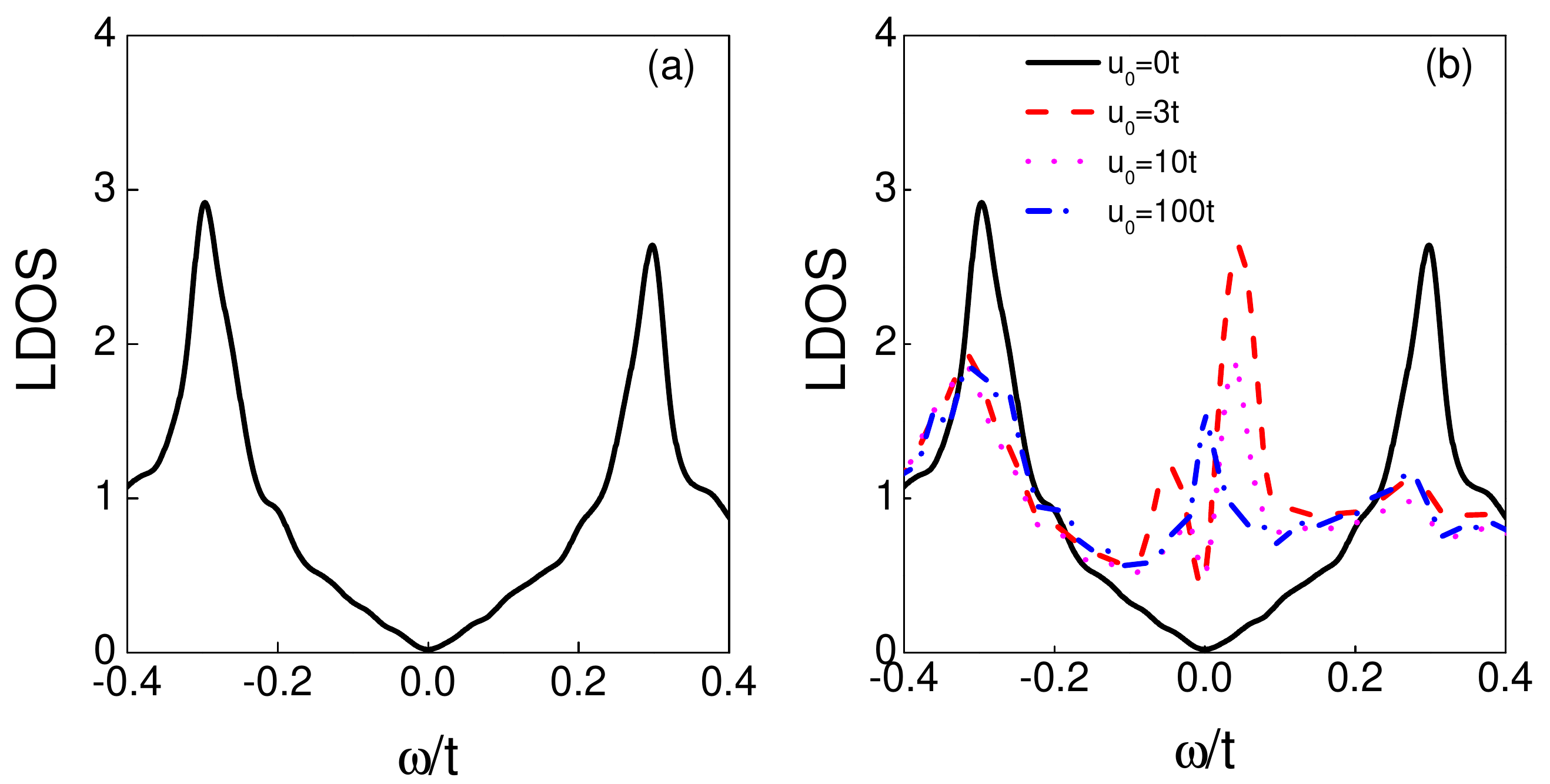}
\end{center}
\caption{(Color online) The LDOS spectra for different scattering potentials $U_{0}$ near the impurity site in the large U limit at $T = 0$.}
\end{figure}

The LDOS spectra for different scattering potentials around the impurity site are plotted in Fig. 2.  For the site which locates far away from the impurity in Fig. 2a, LDOS displays the typical "V"-shaped curve which has recovered the bulk DOS, by exhibiting
a gaplike feature at the gap edges. And especially we find that under the present Gutzwiller-projector
MFA, the LDOS shows the particle-hole asymmetry which have been observed by the STM measurement. At impurity site (not shown here), a single resonance state only appears at small scattering strengthen and is invisible with increasing $U_{0}$ due to the stronger impurity scattering. On the nearest-neighbor site of the impurity (N.N) as seen in Fig. 2b, it is shown that the superconducting coherence peaks are strongly suppressed, and quasiparticle resonance states at intragap energies are generated by a single nonmagnetic impurity. The details features are that for a moderately strong impurity $U_{0}=3t$, the asymmetric resonance states behave to be a double-peaked structure with the $\omega>0$ peak having the dominant spectral weight over the $\omega<0$ peak. While increasing the impurity strength pushes the resonance peaks toward the Fermi level, so that in the unitary limit $U_{0}=100t$, the resonance state occurs right on the Fermi energy, and only a single zero-energy resonance peak appears in the LDOS near the impurity. It is also shown that the effect of the impurity is completely localized. To see clearly this point, we plot the spatial variation of the LDOS at $\omega/t = \pm0.02$ in the unitary limit in Fig. 3 where we can see that the impurity-induced resonance state is indeed localized around the impurity. All the present results are consistent with the experimental data\cite{Yazdani,pan,pan1,Terashima,Zeljkovic,Parham} and previous theoretical calculations\cite{Balatsky0,Franz,Zhu,Morr,wang,Balatsky1}.

\begin{figure}[tbp]
\begin{center}
\includegraphics[width=0.8\linewidth]{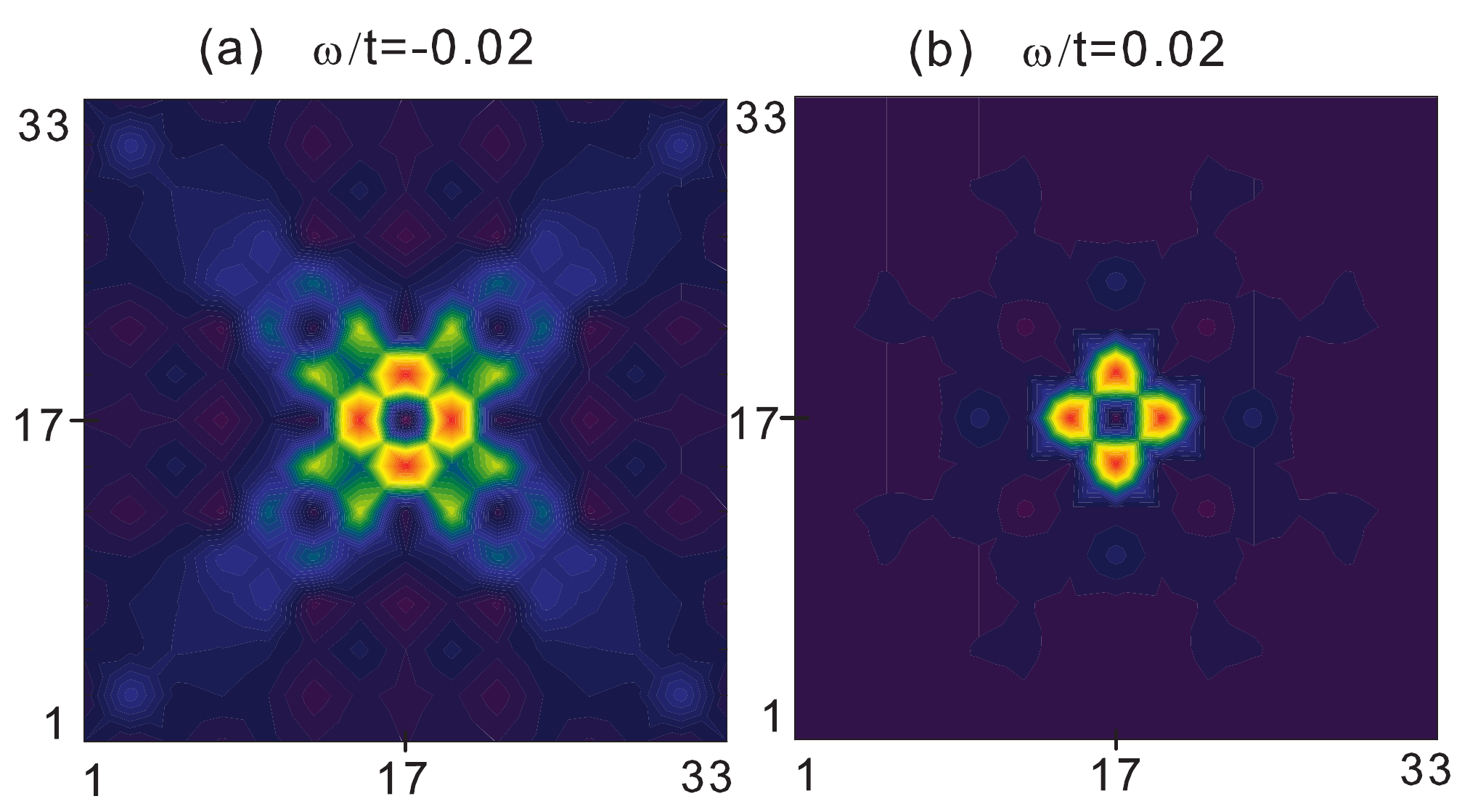}
\end{center}
\caption{(Color online) The spatial variation of the LDOS at $\omega/t = \pm0.02$ in the unitary limit.}
\end{figure}
\begin{figure}[tbp]
\begin{center}
\includegraphics[width=0.9\linewidth]{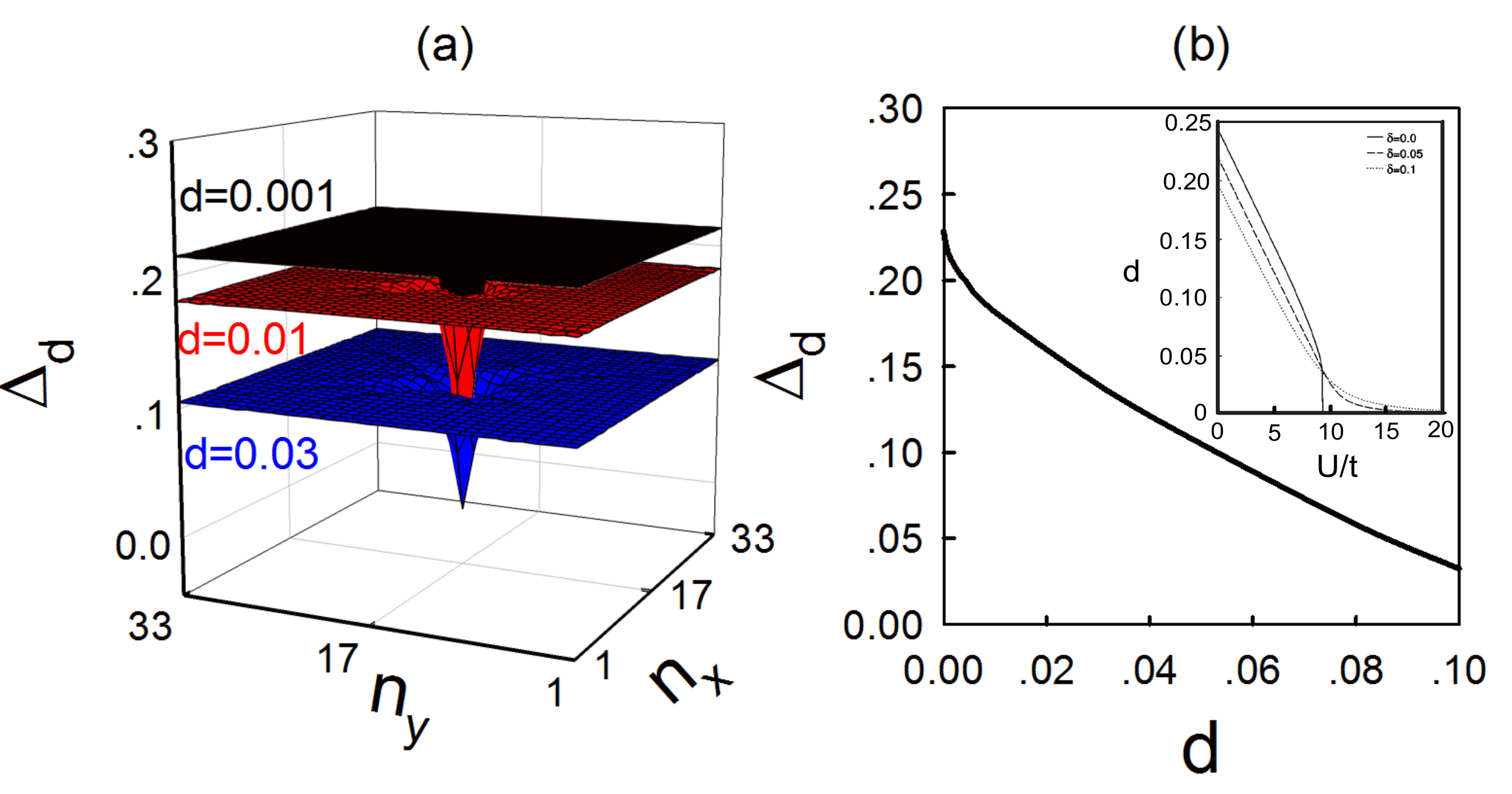}
\end{center}
\caption{(Color online) (a) The spatial variation of the d-wave OP with the EDO $d = 0.001, 0.01, 0.03$ at $T=0$ and $U_{0}=100t$. (b) The d-wave OP as a function of d. Insert: the EDO d as a function of U from Ref. 29.}
\end{figure}

\begin{figure}[tbp]
\begin{center}
\includegraphics[width=0.6\linewidth]{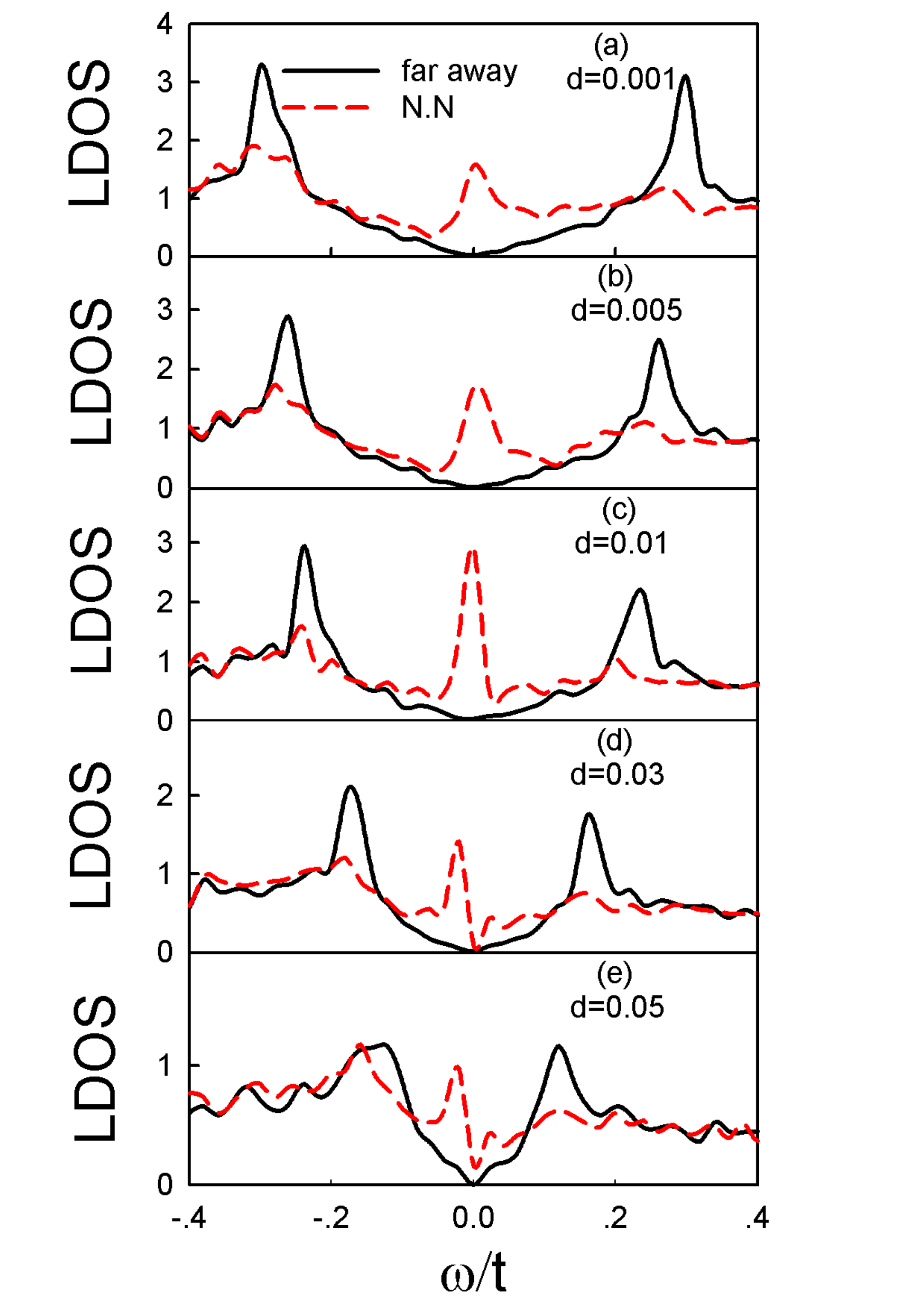}
\end{center}
\caption{(Color online) Evolution of quasiparticle resonance states with the EDO d for the scattering potential $U_{0}=100t$.}
\end{figure}
We now turn to investigate the effect of the Coulomb repulsion U on the quasiparticle resonance states in the impurity scattering unitary limit. The average double occupation number d modulated by U has been studied by one of the authors\cite{Yuan}, where they found that d decreased linearly with increasing Coulomb repulsion U shown as an insert in Fig. 4b. Thus we can directly investigate the effect of the EDO d on the LDOS. The spatial variation of d-wave OP with various $d$ at $T=0$ is self-consistently calculated in Fig. 4a, which indicates that with increasing EDO $d$ (decreasing Coulomb repulsion U), the magnitude of order parameter gradually decreases seen in Fig. 4b. As a result, in Fig. 5 the corresponding position of the superconducting coherent peaks in LDOS move to the lower energies with increasing d, while a single zero-energy impurity resonance peak always survives for small value of d, and begins to evolve into a double-peaked structure with negative energy peak having the dominant spectral weight over the positive energy peak since a critical double occupancy $d_{c}=0.01$. These novel feathers of asymmetric or splitting of the resonance state in the impurity scattering unitary limit near Fermi energy are qualitatively agreement with the STM experiments\cite{Yazdani,pan}. Since the d is modulated by the U, in the large U limit, the no double occupation ($d$=0) constraint is satisfied for the strongly correlated electron systems. As U decreases, the electron double occupation is permitted and the electron correlation becomes weaker, the evolution of quasiparticle resonance states shown above qualitatively describes the effect of the electron correlation interactions on the STM. With the variation of the electron onsite Coulomb interaction, the resonance states induced by the impurity would display different features, which in turn reflects the role played by electron correlation in various cuprate superconductors. In order to avoid the misunderstanding, we stress here again that the evolution of double-peaked resonance into a single one as shown in Fig. 2 just depends on the impurity potential strength $U_{0}$.

\section{Summary}

In conclusion, we have studied the LDOS around a nonmagnetic impurity in the cuprate
superconductors within the Gutzwiller approximation and Bogoliubov-de Gennes theory. We reproduced
the main related experimental results, that is, the asymmetric feature of the LDOS, and impurity induced resonance states which are approximately localized around the impurity. In addition, considering the effect of the Coulomb repulsion, we increase the EDO d which is modulated by U, and find that the OP gradually decreases and the resulting superconducting coherence peaks move to lower energies, while a unitary impurity induced single zero-energy resonance peak always survives for small value of d, and begins to evolve into a double-peaked structure since a critical double occupancy $d_{c}$. These important feathers represent the essential role played by the electron correlations in cuprate superconductors.

\acknowledgments This work was supported by the National Natural Science Foundation of China (NSFC) under Grants No. 11104011, Doctoral Fund of Ministry of Education of China under Grants No. 20110009120024, Research Funds of Beijing Jiaotong University under Grant No. 2013JBM092, and The Project sponsored by SRF for ROCS, SEM.

\end{document}